\documentclass[twocolumn,preprintnumbers,amsmath,amssymb]{revtex4}
\usepackage{graphicx}
\usepackage{amssymb}
\usepackage{rotating} 
\usepackage{epstopdf}  
\usepackage{verbatim}

\newcommand{\be}{\begin{equation}}
\newcommand{\ee}{\end{equation}}
\newcommand{\bea}{\begin{eqnarray}}
\newcommand{\eea}{\end{eqnarray}}
\renewcommand{\l}{\ensuremath{\lambda}}
\newcommand{\lr}{\ensuremath{\text{lr}}}
\newcommand{\sr}{\ensuremath{\text{sr}}}
\renewcommand{\H}{\ensuremath{\text{H}}}
\begin{document}

\title{Range-separated density-functional theory with random phase approximation applied to noncovalent intermolecular interactions}

\author{Wuming Zhu$^{1}$}
\author{Julien Toulouse$^{1}$}
\author{Andreas Savin$^{1}$}
\author{J\'anos G. \'Angy\'an$^2$}

\affiliation{
$^1$Laboratoire de Chimie Th\'eorique, UPMC Univ Paris 06 and CNRS, 75005 Paris, France\\
$^2$CRM2, Institut Jean Barriol, Nancy University and CNRS, 54506 Vandoeuvre-l\`{e}s-Nancy, France
}

 
\begin{abstract}
Range-separated methods combining a short-range density functional with long-range random phase approximations (RPA) with or without exchange response kernel are tested on rare-gas dimers and the S22 benchmark set of weakly interacting complexes of Jure\v cka, \v Sponer, \v Cern\' y, and Hobza ({\it Phys. Chem. Chem. Phys.} {\bf 8}, 1985, 2006). The methods are also compared to full-range RPA approaches. Both range separation and inclusion of the Hartree-Fock exchange kernel largely improve the accuracy of intermolecular interaction energies. The best results are obtained with the method called RSH+RPAx which yields interaction energies for the S22 set with an estimated mean absolute error of about 0.5 - 0.6 kcal/mol, corresponding to a mean absolute percentage error of about 7 - 9\%, depending on the reference interaction energies used. In particular, the RSH+RPAx method is found to be overall more accurate than the range-separated method based on long-range second-order M{\o}ller-Plesset (MP2) perturbation theory (RSH+MP2).
\end{abstract}

\pacs{}
\maketitle

\section{INTRODUCTION}

Because the usual local or semi-local approximations of Kohn-Sham density-functional theory (DFT)~\cite{HohKoh-PR-64,KohSha-PR-65} generally fail to describe non-covalent intermolecular interactions, many approaches have been proposed to remedy their deficiencies. The most widely applied scheme is perhaps the so-called DFT-D approach~\cite{ElsHobFraSuhKax-JCP-01,WuVarNayLotSco-JCP-01,WuYan-JCP-02,ZimParKou-JCP-04,Gri-JCC-04}, in which an empirical dispersion term is added to usual density-functional approximations, using dispersion coefficients generally determined from atomic reference data. There are some efforts to make the DFT-D approach less empirical, for example by calculating dispersion coefficients through the properties of the exchange hole~\cite{BecJoh-JCP-05,Ang-JCP-07}, or from the local response approximation~\cite{SatNak-JCP-09}. Other more or less empirical approaches include using parameterized atom-centered nonlocal one-electron potentials~\cite{LilTavRotSeb-PRL-04} or highly-parameterized hybrid or double-hybrid density functionals (see, e.g., Ref.~\onlinecite{ZhaSchTru-JCTC-06,ChaHea-JCP-09}). Non-empirical approaches include nonlocal correlation functionals derived from response theory~\cite{AndLanLun-PRL-96,DobDin-PRL-96,DioRydSchLan-PRL-04} (possibly combined with long-range corrected exchange functionals~\cite{KamTsuHir-JCP-02,VydVoo-JCP-09,VydVoo-PRL-09}), DFT-based symmetry-adapted intermolecular perturbation theory (see, e.g., Ref.~\onlinecite{HesJanSch-JCP-05}) and range-separated density-functional theory (see, e.g., Ref.~\onlinecite{TouColSav-PRA-04}), in which a short-range density functional is combined with a long-range explicit many-body method such as second-order perturbation theory~\cite{AngGerSavTou-PRA-05}, coupled-cluster theory~\cite{GolWerSto-PCCP-05} or multi-reference second-order perturbation theory~\cite{FroCimJen-PRA-10}.

Recently, the random phase approximation (RPA) and other related approximations to the electron correlation energy have gained revived interest and, in particular, they appear to be viable approaches to the description of weak interactions in molecular and solid-state systems~\cite{Fur-PRB-01,AryMiyTer-PRL-02,FucGon-PRB-02,FucNiqGonBur-JCP-05,FurVoo-JCP-05,MarGarRub-PRL-06,HarKre-PRB-08,Fur-JCP-08,ScuHenSor-JCP-08,RenRinSch-PRB-09,LuLiRocGal-PRL-09,HarKre-PRL-09,NguGir-PRB-09}. However, RPA overestimates short-range correlations between particles~\cite{SinTosLanSjo-PR-68}, giving a too negative correlation energy, which lead Perdew and coworkers~\cite{Per-IJQC-93,KurPer-PRB-99,YanPerKur-PRB-00} to propose the so-called RPA+ scheme in which the RPA correlation energy is corrected by a generalized gradient approximation (GGA). Also, in a Gaussian localized basis, RPA correlation energies have a slow convergence with respect to the basis size~\cite{Fur-PRB-01}. These known flaws of RPA can be cured by the range-separated density-functional theory scheme. Toulouse {\it et al.}~\cite{TouGerJanSavAng-PRL-09} developed a range-separated method combining a long-range RPA-type approximation including the long-range Hartree-Fock exchange kernel with a short-range density functional approximation. Janesko {\it et al.}~\cite{JanHenScu-JCP-09,JanHenScu-JCP-09b,JanScu-JCP-09} have also implemented a range-separated RPA scheme based on the ring-coupled-cluster formulation of RPA~\cite{ScuHenSor-JCP-08}, with no exchange kernel. In their approach, the RPA correlation energy has been rescaled by an empirical coefficient and the range-separation parameter has been optimized in order to improve agreement with experiment.

While initial tests of the range-separated RPA methods for rare-gas dimers and some small molecular complexes show encouraging results \cite{TouGerJanSavAng-PRL-09,JanHenScu-JCP-09,JanHenScu-JCP-09b}, it is not clear how their performance is for weak interactions between larger, biologically important molecules, the main targets for which these hybrid methods were initially developed. In this work, we apply range-separated RPA methods to the S22 test set of Jure\v cka {\it et al.}~\cite{JurSpoCerHob-PCCP-06}. The S22 set comprises 7 hydrogen-bonded systems, 8 dispersion-bonded complexes, and 7 mixed complexes with electrostatic and dispersion interactions. For completeness and comparison, we also include data for homonuclear rare-gas dimers. 

\section{METHODS AND COMPUTATIONAL DETAILS}

The theory of range-separated RPA was described in Ref.~\onlinecite{TouGerJanSavAng-PRL-09}. We give here only a brief overview. In a first step, we perform a self-consistent range-separated hybrid (RSH) calculation~\cite{AngGerSavTou-PRA-05}
\be
E_{\text{RSH}} = \min_{\Phi} \{ \langle \Phi | \hat{T} + \hat{V}_{ne} + \hat{W}^{\lr}_{ee} | \Phi \rangle + E^{\sr}_{\H xc}[n_{\Phi}] \},
\ee
where $\hat{T}$ is the kinetic energy operator, $\hat{V}_{ne}$ is the nuclei-electron potential operator, $\hat{W}_{ee}^{\lr}$ is a long-range electron-electron interaction operator, $E_{\H xc}^{\sr}[n]$ is the associated short-range Hartree-exchange-correlation density functional, and $\Phi$ is a single-determinant wave function. The long-range interaction is constructed with the error function, $w^{\lr}_{ee}(r_{ij}) = \text{erf}(\mu r_{ij})/r_{ij}$, where $\mu$ is the range separation parameter, whose inverse gives the range of the short-range part of the interaction. In this work, we take a fixed value $\mu = 0.5$ bohr$^{-1}$, which seems to be a reasonable choice for a variety of systems~\cite{GerAng-CPL-05a}. Several approximations~\cite{TouSavFla-IJQC-04,TouColSav-PRA-04,TouColSav-JCP-05,GolWerSto-PCCP-05,PazMorGorBac-PRB-06,FroTouJen-JCP-07,GolErnMoeSto-JCP-09} have been proposed for the short-range exchange-correlation functional $E_{xc}^{\sr}[n]$. Here, we use the short-range PBE functional of Ref.~\onlinecite{GolWerSto-PCCP-05}. We have also tested the short-range LDA functional of Ref.~\onlinecite{TouSavFla-IJQC-04}, but since it gives similar results we report here only data obtained with the short-range PBE functional.

The RSH scheme includes long-range Hartree and exchange energies from $\langle \Phi |\hat{W}^{\lr}_{ee} |\Phi \rangle$ as well as short-range Hartree, exchange and correlation energies from the density functional $E^{\sr}_{\H xc}[n]$. The only piece being left out is the long-range correlation energy $E_c^{\lr}$ which needs to be added to the RSH energy
\be
E = E_{\text{RSH}} + E_c^{\lr}.
\ee
The long-range correlation energy $E_c^{\lr}$ can be obtained perturbatively in different ways, for example by second-order M{\o}ller-Plesset (MP2) perturbation theory~\cite{AngGerSavTou-PRA-05,GerAng-CPL-05b,GerAng-JCP-07,GolLeiManMitWerSto-PCCP-08}, coupled-cluster theory~\cite{GolWerSto-PCCP-05,GolWerStoLeiGorSav-CP-06}, RPA-type approximations~\cite{TouGerJanSavAng-PRL-09,JanHenScu-JCP-09}, or multi-reference second-order perturbation theory~\cite{FroCimJen-PRA-10}. In this work, we focus on RPA-type approximations in which $E_c^{\lr}$ can be expressed as an integral over an adiabatic connection~\cite{TouGerJanSavAng-PRL-09}
\be
E_c^{\lr} = \int_{0}^{1} d\lambda \, W^{\lr}_{c,\lambda} 
     = \frac{1}{2} \int_{0}^{1} d\lambda \sum_{iajb} \left < ij | \hat{w}_{ee}^{\lr} | ab  \right > \left( P^{\lr}_{c,\lambda} \right)_{iajb},
\label{Eclr}
\ee
where $i,j$ and $a,b$ refer to occupied and virtual RSH orbitals, respectively, $\left < ij | \hat{w}_{ee}^{\lr} | ab  \right >$ are long-range two-electron integrals, and $P_{c,\lambda}^{\lr}$ is the correlation part of the spin-singlet-adapted two-particle density matrix calculated as~\cite{Fur-PRB-01}
\be
P_{c,\lambda}^{\lr} = 2 \left[ (A_\lambda - B_\lambda)^{1/2} M_\lambda^{-1/2} (A_\lambda - B_\lambda)^{1/2} -1 \right],
\ee
with $M_\lambda = (A_\lambda-B_\lambda)^{1/2} (A_\lambda+B_\lambda) (A_\lambda-B_\lambda)^{1/2}$, and the singlet orbital rotation Hessians $A_\lambda$ and $B_\lambda$. When only the long-range Hartree kernel is included, $A_\lambda$ and $B_\lambda$ write
\be
(A_\lambda)_{iajb} = (\epsilon_{a} - \epsilon_{i}) \delta_{ij} \delta_{ab} + 2 \lambda \left < aj | \hat{w}_{ee}^{\lr} | ib \right >,
\label{HessArpa}
\ee
\be
(B_\lambda)_{iajb} = 2 \lambda \left < ab | \hat{w}_{ee}^{\lr} | ij \right>,
\label{HessBrpa}
\ee
where $\epsilon_{i}$ and $\epsilon_{a}$ are the RSH orbital eigenvalues. We will refer to this method as RSH+RPA (which is equivalent to the method called ``LC-$\omega$LDA+dRPA'' in Refs.~\onlinecite{JanHenScu-JCP-09,JanHenScu-JCP-09b,JanScu-JCP-09} in the special case of the short-range LDA functional). When the long-range Hartree-Fock exchange kernel is also included, $A_\lambda$ and $B_\lambda$ write
\be
(A_\lambda)_{iajb} = (\epsilon_{a} - \epsilon_{i}) \delta_{ij} \delta_{ab} 
+ 2 \lambda \left < aj | \hat{w}_{ee}^{\lr} | ib \right > - \lambda \left < aj | \hat{w}_{ee}^{\lr} | bi \right >,
\label{HessA}
\ee
\be
(B_\lambda)_{iajb} = 2 \lambda \left < ab | \hat{w}_{ee}^{\lr} | ij \right> -  \lambda \left < ab | \hat{w}_{ee}^{\lr} | ji \right >,
\label{HessB}
\ee
and we will refer to this method as RSH+RPAx, as in Ref.~\onlinecite{TouGerJanSavAng-PRL-09}. At second order in the electron-electron interaction, the RSH+RPAx method reduces to the RSH+MP2 method of Ref.~\onlinecite{AngGerSavTou-PRA-05}. The RSH+RPA and RSH+RPAx methods are expected to supersede RSH+MP2 in situations where second-order perturbation theory fails, i.e. when differences of the occupied and virtual orbital energies are small.

The integration over the coupling constant $\lambda$ in Eq.~(\ref{Eclr}) can be performed accurately by a 7-point Gauss-Legendre quadrature~\cite{Fur-PRB-01}, at least for systems not dominated by static correlation effects. In the case of RPA (without exchange-correlation kernel), this integration can also be performed analytically leading to the so-called plasmon formula, as recently emphasized by Furche~\cite{Fur-JCP-08}. The reformulation of RPA as a ring coupled-cluster-doubles theory by Scuseria {\it et al.}~\cite{ScuHenSor-JCP-08}, which is equivalent to the plasmon formula, is another advantageous way of avoiding the numerical integration over $\lambda$. Although there is also a plasmon formula for RPA with the Hartree-Fock exchange kernel which appears in the literature~\cite{MclBal-RMP-64,SzaOst-JCP-77,ScuHenSor-JCP-08}, due to the breaking of antisymmetry of the two-particle density matrix $\left( P^{\lr}_{c,\lambda} \right)_{iajb}$ under exchange of two indices, it does not provide the same correlation energy as Eq.~(\ref{Eclr}) used here and in Ref.~\onlinecite{TouGerJanSavAng-PRL-09}. Unfortunately, as far as we know, the RPAx correlation energy of Eq.~(\ref{Eclr}) cannot be brought to a plasmon formula and the adiabatic connection integral cannot be avoided. However, we show now that it is possible to perform accurately the integration over $\lambda$ by a single-point quadrature. By expanding the integrand in powers of $\l$
\begin{equation}
W_{c,\lambda}^{\lr}= W^{\lr,(1)}_c \lambda + W^{\lr,(2)}_c \lambda^2 + W^{\lr,(3)}_c \lambda^3 + \cdots,
\end{equation}
we can express the long-range correlation energy as the following expansion
\begin{equation}
E_{c}^{\lr} = \int_0^1 d\lambda \, W_{c,\lambda}^{\lr} = \frac{W^{\lr,(1)}_c}{2} + \frac{W^{\lr,(2)}_c}{3} + \frac{W^{\lr,(3)}_c}{4} + \cdots.
\end{equation}
Naively, one could think that a single-point quadrature formula could only be exact up to first order in $\l$. This is the case for example when choosing the quadrature point $\bar{\l}=1$ and weight $1/2$
\begin{equation}
E_{c}^{\lr} \approx \frac{1}{2} W^{\lr}_{c,1},
\end{equation}
or, slightly better, the point $\bar{\l}=1/2$ and weight $1$
\begin{equation}
E_{c}^{\lr} \approx W^{\lr}_{c,1/2}.
\end{equation}
In fact, because the expansion of the integrand starts at first order in $\l$ (i.e., we already know the value of the integrand at the point $\l=0$ which is zero), there is a single-point Radau-type~\cite{AbrSte-BOOK-72} quadrature formula that is exact up to second order in $\l$. Indeed, we find that the condition $w \left( W^{\lr,(1)}_c \bar{\l} + W^{\lr,(2)}_c \bar{\l}^2 \right) = W^{\lr,(1)}_c/2 + W^{\lr,(2)}_c/3$ is fulfilled for the point $\bar{\l}=2/3$ and weight $w=3/4$
\begin{equation}
E_{c}^{\lr} \approx \frac{3}{4} W_{c,2/3}^{\lr}.
\label{twothirdpoint}
\end{equation}
Even better, in the case of the RSH+RPAx method, we can use the long-range MP2 correlation energy which corresponds to the first-order term in $\l$, $E_{c,\text{MP2}}^{\lr}=\int_0^1 d\lambda \, W^{\lr,(1)}_c \lambda = W^{\lr,(1)}_c/2$, in order to find a single-point quadrature formula that is exact up to third order in $\lambda$. Indeed, imposing the condition $w \left(  W^{\lr,(2)}_c \bar{\l}^2 + W^{\lr,(3)}_c \bar{\l}^3 \right) = W^{\lr,(2)}_c/3 + W^{\lr,(3)}_c/4$ leads to the point $\bar{\l}=3/4$ and weight $w=16/27$, so we find
\begin{eqnarray}
E_{c}^{\lr} &=& E_{c,\text{MP2}}^{\lr} + \int_0^1 d\lambda \, \left( W_{c,\lambda}^{\lr} -2 E_{c,\text{MP2}}^{\lr} \, \lambda \right) 
\nonumber\\
&\approx& E_{c,\text{MP2}}^{\lr} + \frac{16}{27} \left( W_{c,3/4}^{\lr} -2 E_{c,\text{MP2}}^{\lr} \, \frac{3}{4} \right) 
\nonumber\\
&=& \frac{1}{9} E_{c,\text{MP2}}^{\lr} + \frac{16}{27} W_{c,3/4}^{\lr}.
\label{threequarterpoint}
\end{eqnarray}
Compared to the 7-point Gauss-Legendre quadrature, the single-point quadrature formulas of Eq.~(\ref{twothirdpoint}) or Eq.~(\ref{threequarterpoint}) significantly reduce the computational cost while introducing only a negligible extra error, as shown later. We will use the formula of Eq.~(\ref{threequarterpoint}) to check our RSH+RPAx results on the S22 set with larger basis sets.

All calculations are done with a development version of the MOLPRO 2008 program~\cite{Molproshort-PROG-08}. The long-range RPA, RPAx and MP2 correlation energies are evaluated with RSH orbitals obtained with the short-range PBE functional of Ref.~\onlinecite{GolWerSto-PCCP-05}. The full-range RPA correlation energy is evaluated with Kohn-Sham orbitals obtained with the usual PBE functional~\cite{PerBurErn-PRL-96}. The full-range RPAx, MP2 and CCSD(T) correlation energies are calculated with Hartree-Fock orbitals. We use the correlation-consistent basis sets of Dunning~\cite{Dun-JCP-89}, or modifications of them. Core electrons are kept frozen in the RPA, RPAx, MP2 and CCSD(T) calculations (i.e. only excitations of valence electrons are considered). Basis set superposition error (BSSE) is removed by the counterpoise method.

For each rare-gas dimer interaction curve, we choose 16 to 19 intermolecular distances, with denser sampling around equilibrium distances. For the S22 set, geometries of the complexes are taken from Ref.~\onlinecite{JurSpoCerHob-PCCP-06} without reoptimization for the different computational methods. The geometries of the isolated monomers are fixed to those in the dimers, thus the so-called monomer deformation energy is not included in the interaction energy. Most of the calculations are done with the same basis sets as those used for the CCSD(T) calculations in Ref.~\onlinecite{JurSpoCerHob-PCCP-06}. The dependence of the interaction energies on the size of basis set is checked by using larger basis sets. For each method, mean error (ME), mean absolute error (MAE) and mean absolute percentage error (MA\%E) are given using as a reference the CCSD(T) values extrapolated to the complete basis set (CBS) limit~\cite{JurSpoCerHob-PCCP-06}. 

\section{RESULTS AND DISCUSSION}

\subsection{Rare-gas dimers}

\begin{figure*} 
\includegraphics[width=18.0cm]{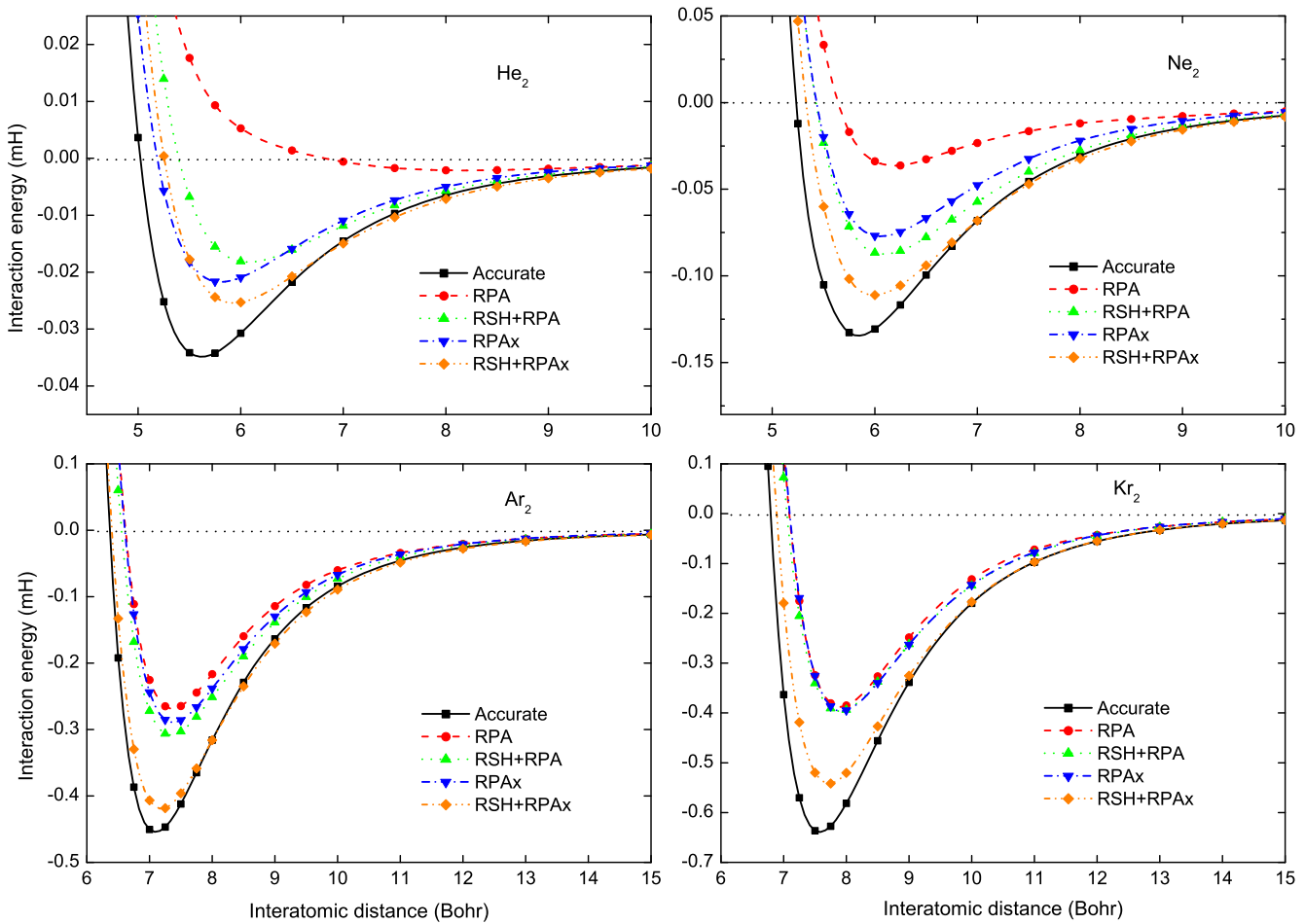}
\caption{Interaction energy curves of rare-gas dimers calculated by the RPA, RSH+RPA, RPAx and RSH+RPAx methods.
Basis sets used are aug-cc-pV5Z. Accurate curves are from Ref.~\onlinecite{TanToe-JCP-03}.}
\label{fig:raregas}
\end{figure*}

We start with rare-gas dimers, which are frequently used for initial tests of methods aiming at describing van der Waals bonded systems. Figure~\ref{fig:raregas} shows the interaction energy curves of He$_2$, Ne$_2$, Ar$_2$, and Kr$_2$ calculated with aug-cc-pV5Z basis sets. The accurate reference curves are from Ref.~\onlinecite{TanToe-JCP-03}. Binding energies at the equilibrium distances obtained with different methods are listed in Table~\ref{tab:raregas}. We can see that full-range RPA can hardly bind two He atoms, and also largely underestimates the binding energy of Ne$_2$. Full-range RPAx and RSH+RPA recover more than half of binding energies for these two systems. RSH+RPAx significantly further improves the results. For larger dimers, Ar$_2$ and Kr$_2$, full-range RPA, full-range RPAx and RSH+RPA all give similar underestimated interaction energies, whereas RSH+RPAx is clearly more accurate. This suggests that range separation and inclusion of the exchange kernel in Eqs.~(\ref{HessA}) and~(\ref{HessB}) are both important. For these systems, RSH+MP2 gives overall similarly accurate equilibrium binding energies than RSH+RPAx.

\begin{table*}[t]
\caption{Interaction energies (in mH) of rare-gas dimers at the equilibrium distance from different methods with aug-cc-pV5Z basis set. For RPA-type methods, the adiabatic connection integration is done by a 7-point Gauss-Legendre quadrature. Mean absolute percentage errors (MA\%E) are given.
} 
\label{tab:raregas}
\begin{tabular}{lccccr} \hline \hline
Method        &     He$_2$ &   Ne$_2$  &  Ar$_2$   &   Kr$_2$  & MA\%E      \\ \hline 
RPA           &     -0.0021 &    -0.0366 &     -0.269 &     -0.388 &      62     \\
RSH+RPA       &     -0.0183 &    -0.0876 &     -0.308 &     -0.397 &      38     \\
RPAx          &     -0.0218 &    -0.0773 &     -0.289 &     -0.396 &      39     \\
RSH+RPAx      &     -0.0255 &    -0.1111 &     -0.420 &     -0.542 &      17     \\
MP2           &     -0.0208 &    -0.0790 &     -0.483 &     -0.691 &      24     \\
RSH+MP2       &     -0.0202 &    -0.1024 &     -0.484 &     -0.671 &      19     \\
CCSD(T)       &     -0.0313 &    -0.1179 &     -0.414 &     -0.575 &      10     \\ \hline
Accurate$^a$  &     -0.0349 &    -0.1342 &     -0.455 &     -0.639 &       0     \\ \hline \hline
\multicolumn{6}{l}{$^a$Ref.~\onlinecite{TanToe-JCP-03}}\\
\end{tabular}
\end{table*}

\subsection{S22 test set}

\begingroup
\squeezetable
\begin{table*}[t]
\caption{Interaction energies (in kcal/mol) for the complexes of the S22 set from different methods with the relatively small basis sets used in Ref.~\onlinecite{JurSpoCerHob-PCCP-06}. For the RPA-type methods, the adiabatic connection integration is done by a 7-point Gauss-Legendre quadrature. The geometries of complexes and the reference CCSD(T)/CBS estimates in the rightmost column are taken from Ref.~\onlinecite{JurSpoCerHob-PCCP-06}. Mean errors (ME), mean absolute errors (MAE) and mean absolute percentage errors (MA\%E) are given.
}
\label{tab:s22}
\begin{tabular}{llrrrrrrrrrrrr} \hline \hline
Complex                           & Basis set$^a$&\hspace{2em}RPA  & RSH+RPA&\hspace{2em}RPAx&RSH+RPAx&\hspace{2em}MP2& RSH+MP2&CCSD(T)&  CCSD(T)/CBS  \\ \hline 
Hydrogen-bonded complexes (HB7)    &          &        &        &          &         &        &          &        &                    \\
1   (NH$_3$)$_2$                   & VQZ      &  -2.31 &  -2.99 &   -2.72  &   -3.19 &  -3.00 &   -3.24  &  -2.96 &    -3.17\\
2   (H$_2$O)$_2$                   & VQZ      &  -3.81 &  -5.21 &   -4.63  &   -5.38 &  -4.71 &   -5.42  &  -4.71 &    -5.02\\
3   Formic acid dimer              & VTZ      & -14.94 & -20.52 &  -17.50  &  -21.00 & -16.89 &  -21.36  & -16.90 &   -18.61\\
4   Formamide dimer                & VTZ      & -12.85 & -16.42 &  -14.51  &  -16.89 & -14.25 &  -17.27  & -14.36 &   -15.96\\
5   Uracil dimer $C_{2h}$          & VTZ-fd   & -15.70 & -20.66 &  -18.28  &  -21.37 & -17.86 &  -22.13  & -17.90 &   -20.65\\
6   2-pyridoxine.2-aminopyridine   & VTZ-fd   & -12.82 & -16.70 &  -14.03  &  -17.39 & -15.08 &  -18.37  & -14.42 &   -16.71\\
7   Adenine.thymine WC             & VDZ      & -10.73 & -15.51 &  -12.39  &  -16.08 & -12.66 &  -16.79  & -12.49 &   -16.37\\
    ME                             &          &   3.33 &  -0.22 &    1.78  &   -0.69 &   1.72 &   -1.16  &   1.82 &    0.00\\
    MAE                            &          &   3.33 &   0.52 &    1.78  &    0.77 &   1.72 &    1.16  &   1.82 &    0.00\\
    MA\%E                          &          & 24.6\% &  4.0\% &  12.7\%  &   5.1\% & 11.0\% &   7.6\%  & 11.8\% &    0.0\%  \\ \hline
\multicolumn{4}{l}{Complexes with predominant dispersion contribution (WI8)}	&        & 	       &		&          & \\
8   (CH$_4$)$_2$                   & VQZ      &  -0.29 &  -0.29 &   -0.28  &   -0.41 &  -0.42 &   -0.45  &  -0.44 &    -0.53\\
9   (C$_2$H$_4$)$_2$               & VQZ      &  -0.92 &  -1.03 &   -1.00  &   -1.35 &  -1.42 &   -1.52  &  -1.31 &    -1.51\\
10  Benzene.CH$_4$                 & VTZ-fd   &  -0.58 &  -0.87 &   -0.58  &   -1.14 &  -1.27 &   -1.51  &  -0.91 &    -1.50\\
11  Benzene dimer $C_{2h}$         & aVDZ     &  -1.35 &  -1.27 &   -0.81  &   -2.04 &  -4.25 &   -4.08  &  -2.03 &    -2.73\\
12  Pyrazine dimer                 & VTZ-fd   &  -1.64 &  -2.46 &   -1.40  &   -3.17 &  -4.94 &   -5.25  &  -2.46 &    -4.42\\
13  Uracil dimer $C_2$             & VTZ-fd   &  -6.16 &  -7.63 &   -5.88  &   -8.67 &  -8.52 &  -10.86  &  -7.24 &   -10.12\\
14  Indole.benzene                 & VDZ      &   0.01 &  -0.94 &    0.56  &   -1.73 &  -3.46 &   -4.30  &  -0.56 &    -5.22\\
15  Adenine.thymine stack          & VDZ      &  -4.21 &  -6.69 &   -4.04  &   -7.84 &  -8.10 &  -11.03  &  -5.40 &   -12.23\\
    ME                             &          &   2.89 &   2.14 &    3.10  &    1.49 &   0.74 &   -0.09  &   2.24 &     0.00 \\
    MAE                            &          &   2.89 &   2.14 &    3.10  &    1.49 &   1.25 &    0.64  &   2.24 &     0.00 \\
    MA\%E                          &          & 57.9\% & 46.3\% &   62.4\% &   28.4\%& 24.2\% &   14.9\% & 39.2\% &     0.0\%   \\ \hline
Mixed complexes (MI7)              &          &        &        &          &         &        &          &        &         \\
16  Ethene.ethine                  & VTZ      &  -0.93 &  -1.31 &   -1.24  &   -1.47 &  -1.43 &   -1.60  &  -1.24 &    -1.53\\
17  Benzene.H$_2$O                 & VTZ-fd   &  -1.97 &  -2.87 &   -2.35  &   -3.10 &  -2.81 &   -3.41  &  -2.48 &    -3.28\\
18  Benzene.NH$_3$                 & VTZ-fd   &  -1.24 &  -1.76 &   -1.39  &   -2.01 &  -1.99 &   -2.36  &  -1.63 &    -2.35\\
19  Benzene.HCN                    & VTZ-fd   &  -2.99 &  -4.39 &   -3.78  &   -4.71 &  -4.41 &   -5.28  &  -3.71 &    -4.46\\
20  Benzene dimer $C_{2v}$         & aVDZ     &  -1.71 &  -1.92 &   -1.65  &   -2.39 &  -3.10 &   -3.33  &  -2.21 &    -2.74\\
21  Indole.benzene T-shape         & VDZ      &  -2.57 &  -3.96 &   -2.90  &   -4.44 &  -4.51 &   -5.50  &  -3.21 &    -5.73\\
22  Phenol dimer                   & VTZ-fd   &  -4.47 &  -6.28 &   -4.99  &   -6.81 &  -6.31 &   -7.72  &  -5.60 &    -7.05\\
    ME                             &          &   1.61 &   0.66 &    1.27  &    0.32 &   0.37 &   -0.29  &   1.01 &     0.00 \\
    MAE                            &          &   1.61 &   0.66 &    1.27  &    0.39 &   0.47 &    0.36  &   1.01 &     0.00 \\
    MA\%E                          &          & 41.2\% & 17.9\% &  31.8\%  &   9.8\% & 11.8\% &   8.9\%  & 24.9\% &    0.0\%  \\ \hline
    total ME                       &          &   2.62 &   0.92 &    2.10  &    0.42 &   0.93 &   -0.49  &   1.71 &     0.00  \\     
    total MAE                      &          &   2.62 &   1.15 &    2.10  &    0.91 &   1.15 &    0.72  &   1.71 &     0.00  \\   
    total MA\%E                    &          & 42.0\% & 23.8\% &  36.9\%  &  15.1\% & 16.0\% &  10.7\%  & 25.9\% &    0.0\%  \\ \hline \hline
\multicolumn{14}{l}{$^a$VDZ, aVDZ, VTZ, and VQZ stand for cc-pVDZ, aug-cc-pVDZ, cc-pVTZ, and cc-pVQZ, respectively. In the modified VTZ-fd basis set, the set}\\
\multicolumn{14}{l}{of f functions and the tight d functions are removed from cc-pVTZ basis (for hydrogen the set of d functions and the set of tight p functions are}\\
\multicolumn{14}{l}{removed).}\\
\end{tabular}
\end{table*}
\endgroup

Although the rare-gas dimers usually constitute a good initial test for a method describing non-covalent bonding, they are still not sufficient to assess the reliability of a method for applications to biologically interesting systems~\cite{CerHob-PCCP-05}. It is necessary to include a few moderately large molecules in a reliable test set. We have then taken the S22 set of Jure\v cka {\it et al.}~\cite{JurSpoCerHob-PCCP-06}, which composes 22 weakly-bonded molecular complexes, including 7 hydrogen-bonded complexes (HB7 subset), 8 weakly-interacting complexes with predominant dispersion contributions (WI8 subset) and 7 mixed complexes featuring multipole interactions (MI7 subset). We use the same names of the subsets introduced in Ref.~\onlinecite{GolLeiManMitWerSto-PCCP-08}. Recently, Marchetti and Werner did explicitly-correlated coupled-cluster calculations for the S22 set and found close agreement with the original benchmark data~\cite{MarWer-JPCA-09}. When this paper is under review, Takatani {\it et al.}~\cite{TakHohMalMarShe-JCP-10} published a new benchmark for the S22 test set, which is supposed to be more accurate. We still use the original benchmark data of Ref.~\onlinecite{JurSpoCerHob-PCCP-06} for detailed comparisons, but the overall effects of using other reference data are discussed at the end of this section.

\subsubsection{Comparison of all the methods with small basis sets}

\begin{figure} 
\includegraphics[width=9.0cm]{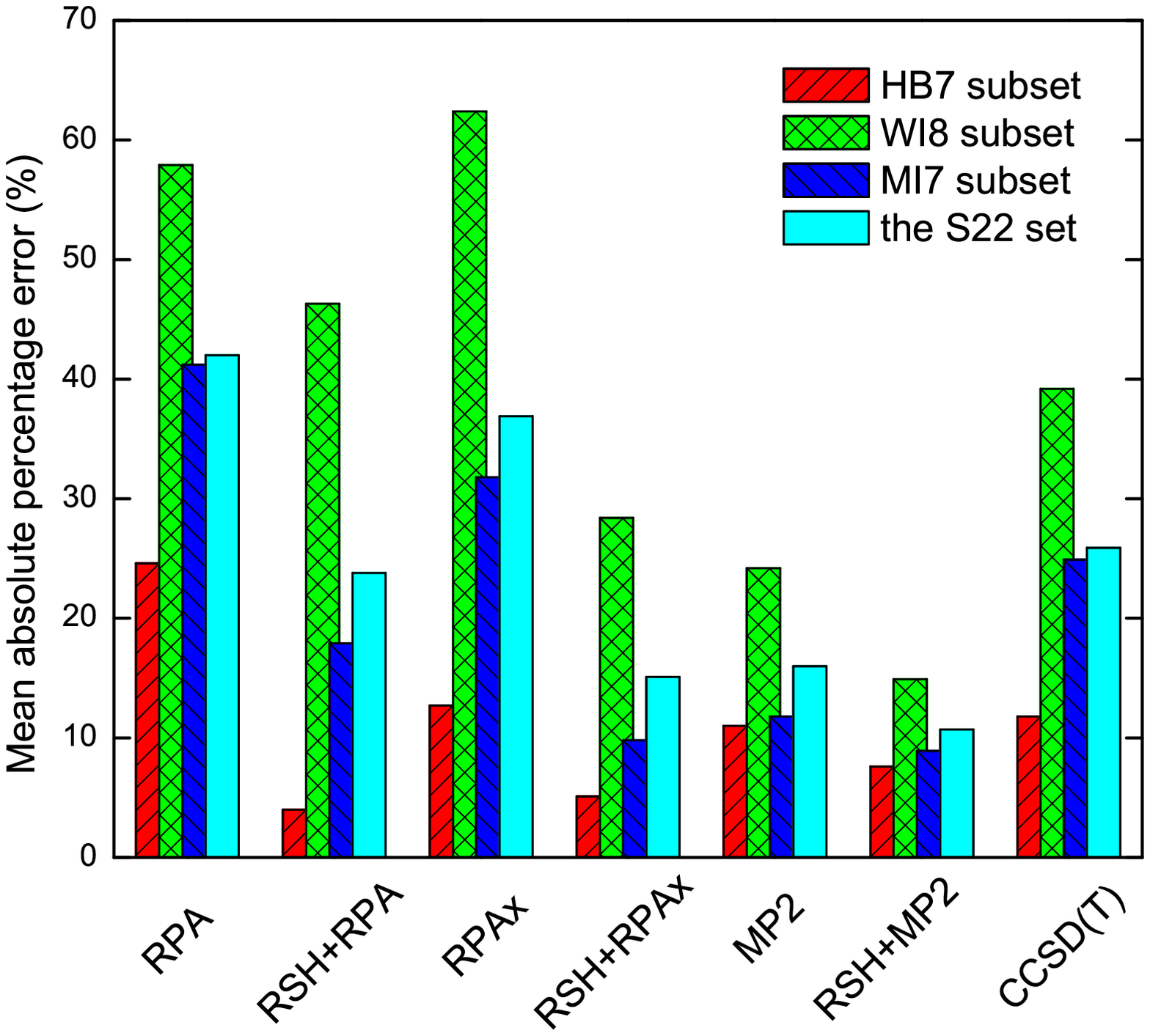}
\caption{Mean absolute percentage errors for the S22 test set and for its three subsets from different methods with the relatively small basis sets used in Ref.~\onlinecite{JurSpoCerHob-PCCP-06}. The data are from Table~\ref{tab:s22}.}
\label{fig:s22}
\end{figure}

Interaction energies calculated with RPA, RSH+RPA, RPAx, RSH+RPAx, MP2, RSH+MP2 and CCSD(T) are given in Table~\ref{tab:s22} with the same relatively small basis sets as those used for the CCSD(T) calculations in Ref~\onlinecite{JurSpoCerHob-PCCP-06}. The basis sets used for the 22 complexes are not uniform because the size of complexes varies a lot. VTZ-fd basis sets are obtained by removing the set of f functions and the set of tight d functions from cc-pVTZ (for hydrogen the set of d functions and the set of tight p functions are removed).
The rightmost column contains CCSD(T)/CBS estimates taken from Ref.~\onlinecite{JurSpoCerHob-PCCP-06} which serve as reference values. The mean absolute percentage error of each method for each subset and for the total S22 set are also shown in Fig.~\ref{fig:s22}.

For the hydrogen-bonded systems (HB7 subset), full-range RPA and RPAx underestimate the interaction energies on average by about 25\% and 13\%, respectively. Range separation greatly improves these two methods, the MA\%E of RSH+RPA and RSH+RPAx on this subset being 4.0\% and 5.1\%, respectively. Note that all the range-separated methods overestimate the interaction energies, whereas all other methods underestimate them.

For the dispersion-bonded systems (WI8 subset), accurate treatment of correlation is crucial. While Hartree-Fock calculations account roughly for two thirds of the binding energy in the complexes of the HB7 subset, it does not predict any bonded complex in the WI8 subset. As expected, including short-range correlation does not help much. Except for stacked uracil dimer, RSH calculations (without long-range correlation) do not predict any bonded complex either. Standard DFT calculations with the PBE functional do give 5 bonded complexes out of 8, but the MA\%E on the WI8 subset is still larger than 100\%. It is clear that a good, physically well-founded description of long-range correlation is absolutely necessary. Full-range RPA recovers nearly half of the interaction energies for the complexes in the WI8 subset, except for the stacked indole-benzene complex for which it gives zero binding. Full-range RPAx is similar in performance but slightly worsens the results. Range separation improves RPA but the remaining mean absolute percentage error (about 46\%) remains quite large. A much more significant improvement is obtained with RSH+RPAx which reduces the MA\%E to about 28\%. The two largest errors from the RSH+RPAx method are for stacked indole-benzene and stacked adenine-thymine, 3.5 kcal/mol and 4.4 kcal/mol, or 67\% and 36\%, respectively. They seem quite large but are still smaller than the basis set errors in CCSD(T), which are 89\% and 56\%, respectively. It turns out that at least half of the error is from using too small basis sets (see next subsection). With these small basis sets, MP2 and RSH+MP2 give more accurate interaction energies than RSH+RPAx with a MA\%E on this subset of about 24\% and 15\%, respectively, but this apparently good performance can be mostly attributed to the basis set error which partially cancels the MP2 overestimation of the binding energies. The errors in MP2 and RSH+MP2 thus tend to increase as the basis set is enlarged. In fact, on the WI8 subset, the MA\%E of MP2 extrapolated to CBS~\cite{JurSpoCerHob-PCCP-06} is 33\%, and the MA\%E of range-separated MP2 (with local correlation and density fitting approximations) with aug-cc-pVTZ basis~\cite{GolLeiManMitWerSto-PCCP-08} is 24\%. On the contrary, the errors in RSH+RPAx tend to decrease when we use larger basis sets, as it will be illustrated in the next section.

For the mixed complexes (MI7), the errors are somewhere in between those for the HB7 and the WI8 subsets. For RSH+RPAx, the largest error is for T-shaped indole-benzene, 1.3 kcal/mol or 23\%. Again the larger part of the error is from incomplete basis set, as it will be shown in the next subsection.

Among the methods that we have investigated here, RSH+RPAx and RSH+MP2 appear to be the most accurate ones. We thus further investigate these two methods with larger basis sets.

\subsubsection{RSH+RPAx and RSH+MP2 results with larger basis sets}

\begingroup
\squeezetable
\begin{table*}[t]
\caption{Interaction energies (in kcal/mol) for the complexes of the S22 set from the RSH+RPAx and RSH+MP2 methods with augmented basis sets. (aVDZ and aVTZ stand for aug-cc-pVDZ and aug-cc-pVTZ, respectively.) The adiabatic connection integration is done by the single-point quadrature of Eq.~(\ref{threequarterpoint}). The geometries of complexes and the reference CCSD(T)/CBS estimates in the rightmost column are taken from Ref.~\onlinecite{JurSpoCerHob-PCCP-06}. Italic numbers for RSH+MP2 with aVTZ basis were taken from Ref.~\onlinecite{GolLeiManMitWerSto-PCCP-08}. Italic numbers for RSH+RPAx with aVTZ basis were estimated by E$_{int}$(RSH+RPAx/aVTZ) = E$_{int}$(RSH+RPAx/aVDZ) + (E$_{int}$(RSH+MP2/aVTZ) - E$_{int}$(RSH+MP2/aVDZ)). Mean errors (ME), mean absolute errors (MAE) and mean absolute percentage errors (MA\%E) are given.
}
\label{tab:s22largerbasis}
\begin{tabular}{lrrrrrrrrrrrr} \hline \hline
                                     &    \multicolumn{2}{c}{RSH+RPAx} &  \multicolumn{2}{c}{RSH+MP2}    &   CCSD(T)/CBS \\ 
Complex                              &    aVDZ     &  aVTZ    &    aVDZ &   aVTZ  &     \\ \hline       
Hydrogen-bonded complexes (HB7)      &             &          &         &         &              \\                
1   (NH$_3$)$_2$                     &    -3.07    &   -3.19  &   -3.13 &   -3.25 &      -3.17\\                   
2   (H$_2$O)$_2$                     &    -5.33    &   -5.41  &   -5.37 &   -5.45 &      -5.02\\                   
3   Formic acid dimer                &   -20.81    &  -21.18  &  -21.20 &  -21.57 &      -18.61\\                  
4   Formamide dimer                  &   -17.03    &  -17.22  &  -17.44 &  -17.64 &      -15.96\\                  
5   Uracil dimer $C_{2h}$            &   -21.80    &  \emph{-22.00}  &  -22.62 &  \emph{-22.82} &     -20.65\\                   
6   2-pyridoxine.2-aminopyridine     &   -17.81    &  \emph{-17.55}  &  -18.86 &  \emph{-18.60} &     -16.71\\                   
7   Adenine.thymine WC               &   -17.29    &  \emph{-17.15}  &  -18.26 &  \emph{-18.12} &     -16.37\\                   
    ME                               &    -0.95    &  -1.03   &   -1.48 &   -1.57 &      0.00 \\                   
    MAE                              &     0.98    &   1.03   &    1.50 &    1.57 &      0.00 \\                   
    MA\%E                            &     6.5\%   &   6.6\%  &    9.3\%&   10.0\%&      0.0\%\\ \hline            
\multicolumn{4}{l}{Complexes with predominant dispersion contribution (WI8)}   &               \\
8   (CH$_4$)$_2$                     &    -0.42    & -0.45    &   -0.46 &   -0.48 &         -0.53\\                
9   (C$_2$H$_4$)$_2$                 &    -1.28    & -1.38    &   -1.45 &   -1.55 &         -1.51\\                
10  Benzene.CH$_4$                   &    -1.23    & -1.32    &   -1.62 &   -1.71 &         -1.50\\                
11  Benzene dimer $C_{2h}$           &    -2.05    & \emph{-2.21}    &   -4.08 &  \emph{-4.24} &         -2.73\\                
12  Pyrazine dimer                   &    -3.78    & \emph{-3.85}    &   -5.97 &  \emph{-6.04} &         -4.42\\                
13  Uracil dimer $C_2$               &    -9.38    & \emph{-9.57}    &  -11.76 &  \emph{-11.59} &        -10.12\\                
14  Indole.benzene                   &    -3.70    & \emph{-3.71}    &   -6.95 &  \emph{-6.96} &         -5.22\\                
15  Adenine.thymine stack            &   -10.97    & \emph{-10.57}   &  -15.11 &  \emph{-14.71} &        -12.23\\                
    ME                               &     0.68    &  0.65    &   -1.14 &   -1.13 &      0.00 \\                   
    MAE                              &     0.68    &  0.65    &    1.18 &    1.14 &      0.00 \\                   
    MA\%E                            &    17.4\%   & 14.4\%   &   22.8\%&   23.3\%&      0.0\%\\ \hline            
Mixed complexes (MI7)                &             &               &     &               &            &      \\    
16  Ethene.ethine                    &    -1.48    &    -1.54 &   -1.62 &   -1.68 &          -1.53\\               
17  Benzene.H$_2$O                   &    -3.16    &    -3.33 &   -3.49 &   -3.68 &          -3.28\\               
18  Benzene.NH$_3$                   &    -2.11    &    -2.24 &   -2.49 &   -2.63 &          -2.35\\               
19  Benzene.HCN                      &    -4.54    &    -4.77 &   -5.13 &   -5.38 &          -4.46\\               
20  Benzene dimer $C_{2v}$           &    -2.39    &    \emph{-2.55} &   -3.33 &   \emph{-3.49} &          -2.74\\               
21  Indole.benzene T-shape           &    -5.17    &    \emph{-5.46} &   -6.55 &   \emph{-6.84} &          -5.73\\               
22  Phenol dimer                     &    -7.07    &    \emph{-7.11} &   -8.05 &   \emph{-8.09} &       -7.05\\                  
    ME                               &     0.17    &     0.02 &   -0.50 &   -0.66 &       0.00 \\                  
    MAE                              &     0.20    &     0.14 &    0.50 &    0.66 &       0.00 \\                  
    MA\%E                            &     6.0\%   &     3.8\%&   11.9\%&   16.6\%&       0.0\%\\ \hline           
    total ME                         &     0.00    &    -0.08 &   -1.05 &   -1.12 &     0.00 \\                    
    total MAE                        &     0.62    &     0.61 &    1.06 &    1.12 &         0.00 \\                
    total MA\%E                      &    10.3\%   &     8.6\%&   15.1\%&   16.9\%&     0.0\%\\\hline \hline          
\end{tabular}
\end{table*}
\endgroup

For the largest systems in Table~\ref{tab:s22}, the stacked indole-benzene and stacked adenine-thymine complexes, only the cc-pVDZ basis set was used and this corresponds to the two largest errors in RSH+RPAx. Although range-separated methods converge much faster with respect to basis size than full-range methods~\cite{TouGerJanSavAng-PRL-09}, the cc-pVDZ basis set is so small that we may suspect that those errors are mainly a consequence of the incompleteness of the basis. We did RSH+RPAx and RSH+MP2 calculations with larger augmented basis sets, aug-cc-pVDZ and aug-cc-pVTZ, using the approximate single-point quadrature of Eq.~(\ref{threequarterpoint}) for RSH+RPAx to keep the computational cost acceptable, and the interaction energies are given in Table~\ref{tab:s22largerbasis}. For the aug-cc-pVTZ basis set, as the calculations are expensive, we use for the largest systems the RSH+MP2 interaction energies with local correlation and density fitting approximations calculated by Goll {\it et al.}~\cite{GolLeiManMitWerSto-PCCP-08} and estimate the RSH+RPAx interaction energies from the aug-cc-pVDZ calculations by the simple correction formula: E$_{int}$(RSH+RPAx/aVTZ) = E$_{int}$(RSH+RPAx/aVDZ) + (E$_{int}$(RSH+MP2/aVTZ) - E$_{int}$(RSH+MP2/aVDZ)). This correction can be checked for the complexes for which we have the actual aug-cc-pVTZ calculations, and it appears to work well. 

We have also checked that the single-point quadrature approximation does not introduce any significant error. By recalculating the RSH+RPAx interaction energies of the S22 set with the single-point quadrature of Eq.~(\ref{threequarterpoint}) with the same basis sets used in Table~\ref{tab:s22} and comparing to the interaction energies of Table~\ref{tab:s22}, which were obtained from a 7-point Gauss-Legendre quadrature, we find a mean error of -0.004 kcal/mol, or 0.11\%. The largest difference is for stacked indole-benzene, which is -0.014 kcal/mol, or 0.81\%, still significantly smaller than the error in the method itself. For comparison, the less accurate other single-point quadrature of Eq.~(\ref{twothirdpoint}) gives a mean error of +0.07 kcal/mol, or 0.5\%, still a very good approximation.

Comparison of Tables~\ref{tab:s22} and~\ref{tab:s22largerbasis} shows that the interaction energies of the complexes in the WI8 subset calculated with the cc-pVDZ or VTZ-fd basis are seriously underestimated. Note that the aug-cc-pVDZ basis is larger than the VTZ-fd basis and expectedly appears to be of a better quality. From Table~\ref{tab:s22} to Table~\ref{tab:s22largerbasis}, the errors for the complexes number 12 to 15 are reduced by about a factor of two or more. Correspondingly, the MA\%E of RSH+RPAx on the WI8 subset decreases from about 28\% to 14\% with aug-cc-pVTZ basis. On the contrary, the MA\%E of RSH+MP2 increases from 15\% to 23\%, confirming that the good value in Table~\ref{tab:s22} was due to a fortuitous compensation of errors. Even for range-separated methods, including diffuse functions in the basis sets for dispersion-bonded systems is essential, in order to ensure a reasonable description of monomer polarizabilities.

The evolution from Table~\ref{tab:s22} to Table~\ref{tab:s22largerbasis} is similar for the MI7 subset. For T-shaped indole-benzene, the error in RSH+RPAx is reduced from 1.3 kcal/mol to 0.3 kcal/mol, or from 23\% to 5\%. On this subset, the MA\%E of RSH+RPAx decreases from 10\% to 3.8\% with aug-cc-pVTZ basis while the MA\%E of RSH+MP2 increases from 9\% to 17\%.

The hydrogen-bonded systems are less sensitive to basis sets than the dispersion-bonded systems. The overestimation of the interaction energies by RSH+RPAx and RSH+MP2 for the HB7 subset is a bit reinforced with the augmented basis sets of Table~\ref{tab:s22largerbasis}. The most serious overestimation is for the formic acid dimer, -2.6 kcal/mol and -3.0 kcal/mol with aug-cc-pVTZ for RSH+RPAx and RSH+MP2, respectively. The error is most likely due to a deficiency in the approximate short-range density functional, because the RSH calculation (without long-range correlation) predicts an interaction energy (-18.54 kcal/mol) already very close to the reference value (-18.61 kcal/mol). This conjecture is also supported by the comparison between the MP2 and RSH+MP2 columns in Table~\ref{tab:s22}. MP2 systematically underestimates whereas RSH+MP2 consistently overestimates interactions in the HB7 subset, reflecting an overbinding feature of the short-range functional that we used. This point has already been remarked by Goll {\it et al.}~\cite{GolLeiManMitWerSto-PCCP-08}.
 
Overall, with aug-cc-pVTZ basis, RSH+RPAx yields interaction energies for the total S22 set with an estimated MAE of 0.61 kcal/mol or MA\%E of 8.6\%, while RSH+MP2 gives a MAE of 1.12 kcal/mol or MA\%E of 16.9\%. Using the reference data by Marchetti and Werner~\cite{MarWer-JPCA-09} would change these MAEs by less than 0.06 kcal/mol and corresponding MA\%Es by less than 1\%. If we use the latest benchmark from Takatani {\it et al.}~\cite{TakHohMalMarShe-JCP-10}, the MAE and MA\%E for RSH+RPAx reduce to 0.46 kcal/mol and 6.7\%, respectively, whereas the MAE and MA\%E for RSH+MP2 increase to 1.17 kcal/mol and 18.9\%, respectively.

\section{SUMMARY AND CONCLUSIONS}

We have tested and compared four RPA-based methods, namely the full-range RPA and RPAx methods, and the range-separated RSH+RPA and RSH+RPAx methods, on homonuclear rare-gas dimers and the S22 set of weakly-interacting intermolecular complexes of Jure\v cka {\it et al.}~\cite{JurSpoCerHob-PCCP-06}. Both range separation and inclusion of the Hartree-Fock exchange response kernel largely improve the accuracy of the predicted interaction energies. The best method, RSH+RPAx, gives satisfactory interaction energy curves of the rare-gas dimers and yields interaction energies of the S22 set with an estimated mean absolute error of about 0.5 - 0.6 kcal/mol, corresponding to a mean absolute percentage error of about 7 - 9\% (depending on the reference interaction energies used), with adequate basis sets including diffuse functions. The RSH+RPAx method is found to be overall more accurate than the simpler RSH+MP2 method, although the latter remains a very reasonable approach for weak intermolecular interactions.

As RSH+RPAx still seems to underestimate systematically interaction energies for stacked complexes and to overestimate the strength of hydrogen bonds, further refinements to this approach are desired. One may improve the short-range density functional, for example by using a meta-GGA form~\cite{GolErnMoeSto-JCP-09}. One may improve the long-range correlation energy by using other variants or extensions of RPA-type approximations~\cite{PaiJanHenScuGruKre-JCP-10,GamCatGra-PRC-09}. One may adjust the value of the separation parameter which was fixed to $\mu = 0.5$ bohr$^{-1}$ in this study. Finally, one may refine the form the decomposition of the electron-electron interaction~\cite{TouColSav-PRA-04,YanTewHan-CPL-04,HenIzmScuSav-JCP-07,SonWatHir-JCP-09}.

Although our current implementation remains expensive compared to standard DFT calculations, we have shown that the computational cost of the adiabatic connection integration can be reduced by using an approximate single-point quadrature [Eq.~(\ref{twothirdpoint}) or Eq.~(\ref{threequarterpoint})], without any significant loss of precision. In view of the recent progress in the development of efficient RPA algorithms~\cite{Fur-JCP-08,ScuHenSor-JCP-08}, further computational improvements of the RSH+RPAx method or other closely-related variants can be expected. Nevertheless, the tests performed with our current implementation have already demonstrated that RSH+RPAx is a feasible method and provides a reasonable description to non-covalent bonding in medium to large sized complexes, including biomolecules.

{\it Note added in proof:} While proofreading the manuscript, a paper has been published online, Podeszwa {\it et al.}, Phys. Chem. Chem. Phys., DOI:10.1039/b926808a (2010), reporting an improved set of S22 reference interaction energies. Using these data the MAE and MA\%E for RSH+RPAx become 0.49 kcal/mol and 7.0\%, and for RSH+MP2, 1.19 kcal/mol and 19.1\%, respectively. These deviations are very similar to those found with respect to the data of Ref. 66.

\section*{ACKNOWLEDGMENTS}

We thank Peter Reinhardt and Hermann Stoll for helpful discussion and technical assistance. This work is financed through ANR (French national research agency) via contract number ANR-07-BLAN-0272 (Wademecom).



\end{document}